\documentstyle[twoside,fleqn]{article}                               
\newcounter{zacountsec}   
\newcommand{\eh}{\hfill}\newlength{\sperr}                                    
\newcommand{\Title}[1]{{\large \bf #1}}
\newcommand{\Author}[1]{{\sc #1}}
\newcommand{\Section}[1]{{\stepcounter{zacountsec}\vspace{3mm}%
\hspace*{12mm}\normalsize\bf\arabic{zacountsec}. \parbox[t]{150mm}{ #1 }}}
                     
\newenvironment{Abstract}{\begin{minipage}[t]{177mm}\em }%
{\end{minipage}}
\newenvironment{thmm}[2]{\begin{sloppypar}%
{#1 #2.}\em{}}%
{\end{sloppypar}}

\newcommand{\proof}{\hspace*{9mm}{\settowidth{\sperr}{\rm Proof}%
\parbox[t]{1.3\sperr}{\rm P\eh r\eh o\eh o\eh f\eh. } }}

\newcounter{zalit}

\newenvironment{References}{%
\Section{References}%
\begin{small}\begin{list}{\arabic{zalit} }{\usecounter{zalit}    
\itemsep0mm \parsep0mm\settowidth{\labelwidth}{\small\rm 88}\labelsep0mm
\setlength{\leftmargin}{\labelwidth}}}%
{\end{list}\end{small}}
\newlength{\addro}\newlength{\addrt}                                   
\newenvironment{Address}{{\em Address: }%
\begin{minipage}[t]{\addrt}}%
{\end{minipage}}

\makeatother
\newcommand{\EQ}{\begin{equation}}
\newcommand{\EN}{\end{equation}}
\newcommand{\BEQ}{{\vskip-.05in} \begin{equation}}
\newcommand{\NEQ}{\end{equation} {\vskip-.05in} }
\def\ni{\noindent}

\def\Ah{\hat{A}}
\def\omh{\hat{\omega}}

\def\part{\partial}

\begin{document}
%\vspace*{15mm}\hspace*{18mm}
\vspace*{8mm}\hspace*{18mm}
%{\vskip-.1in}
\begin{minipage}[t]{157mm}     
%****************************************************************************
%****************************************************************************
%******************* Please insert here the name(s) of the author(s). *******
\Author{Kurt S.~Riedel}
\vspace*{0.35cm} 

\Title{Signal Processing and
Piecewise Convex Estimation}
\end{minipage}
\vspace*{2mm}      

\begin{Abstract} 
Many problems on signal processing reduce to nonparametric function estimation.
%Three standard techniques are smoothing splines, kernel
%smoothers and wavelet thresholding; all of which have data adaptive
%versions. 
We propose a new methodology, piecewise convex fitting (PCF), 
and give a two-stage adaptive estimate.
%A two-stage estimator is proposed and shown to achieve
%the optimal rate of convergence of the MSE.    
In the first stage, the number and location of the change points
is estimated using strong smoothing. In the second stage, a 
constrained smoothing
spline fit is performed with the smoothing level  chosen to minimize the
MSE. 
The imposed constraint is that a single  change point
occurs in a region about each empirical change point of the 
first-stage estimate. This constraint is equivalent to requiring
that the third derivative of the second-stage estimate has a
single sign in a small neighborhood about each first-stage
change point.
We sketch how PCF may be applied to signal recovery, instantaneous frequency
estimation, surface reconstruction, image segmentation, spectral estimation
and multivariate adaptive regression.
\end{Abstract}

%{\vskip-.40in}

\Section{Signal Recovery}

Techniques such as  splines and wavelets are optimal for estimating an 
arbitrary function in a ball in function space. We claim that the
``uniform'' prior is unrealistic and that
 naturally occuring functions have very few inflection points. 
%the estimated function should have as few inflection points as possible. 
 Our basic tenet is  that the fitted curve should preserve the geometric
fidelity of the unknown function by having the same number of inflections.
%The resulting
%``geometrically accurate'' fit will have a slightly larger mean square
%error (MSE) an arbitrary function in a ball in function space. We conjecture
%that our methods will result in a lower MSE for functions with only a few
%inflection points.
Consider a signal, $y_i = g(t_i )+ \epsilon_i$, measured at $t_i =
i\delta$, $i=1\ldots N$, where $\{\epsilon_i\}$ are independent Gaussian random
variables: $\epsilon_i \sim N(0,\sigma^2 )$. 
Let $g(t) $ have $K$ change points of $\ell$-convexity  with change points
$x_1 \leq x_2 \ldots \leq x_K$ if $(-1)^{k-1} g^{(\ell )} (t) \geq 0$ for
$x_k \leq t \leq x_{k+1}$.  In practice, we take $\ell =2$. 

Our goal is to {\em estimate $g(t)$ while the preserving the geometry of $g$.} 
A standard technique, kernel smoothers [10], estimates $g(t)$ by a
weighted local average:
$
\hat{g}_h (t) = {1\over Nh} \sum_{i=1}^N y_i \kappa \left(
{t-t_i \over h } \right) 
$,
where $h$ is the kernel halfwidth that determines the amount of
smoothing. As $h$ increases, the random error (variance) in
$\hat{g}$ decreases, while the systematic error (bias) increases. 
%Recent data adaptive schemes choose a $t$ dependent halfwidth
%$h(t)$ to minimize an empirical estimate of the local MSE.
In [8], the geometric faithfulness of kernel smoothers is examined in the
limit as $N\rightarrow\infty$ and $h\rightarrow 0$ with $N\delta =1$.
The halfwidth that minimizes the mean square error scales as
$N^{-1\over 2 m+1}$ for $g\in C^m [0,1]$ provided that the kernel
satisfies certain moment conditions. 
For $\ell =m$ or $ m-1$, this halfwidth scaling,
$N^{{-1\over 2m+1}}$, produces extra (artificial) $\ell $-change points
with high probability. To eliminate the artificial inflection points with
probability 1, the smoothing must be increased such that $h >>
N^{{-1\over 2\ell+3} }ln[N]$. %\rightarrow \infty$. 
Thus, the  level of smoothing required for geometric fidelity is large
enough to degrade the MSE.
In [8], we propose a two-stage estimator.

\ni
Stage 1: Strongly smooth with $h_N N^{{1\over 2\ell +3}} / \ln N$
$\rightarrow\infty$. Denote the empirical $\ell$-change points by
$\hat{x}_{k} ,k =1\ldots\hat{K}$.

\ni
Stage 2: Perform a constrained smoothing spine fit by minimizing 
the penalized likelihood
subject to the constraints that $\hat{g}^{(\ell +1)} (t)$ does not
change sign in the intervals, 
$[\hat{x}_k -z_{\alpha} \hat{\sigma}_k , \hat{x}_k +z_{\alpha}
\hat{\sigma}_k ]$.
The $k$-th empirical change point variance is % $\hat{\sigma}_k$ is
$\hat{\sigma}_k^2 \equiv \left[ {c\sigma^2 \over
| \hat{g}_{stage \ 1}^{(\ell +1)} (\hat{x}_k )|^2 Nh^{2\ell +3} }
\right]$,
where $\hat{g}_{stage \ 1}^{(\ell +1)} (\hat{x}_k )$ is the estimate
of $g^{(\ell +1)}$
at $\hat{x}_k$ from the first stage. $c$ is a
constant that depends on the kernel shape. The confidence interval
parameter, $z_{\alpha}$, is the $\alpha$-quantile for the normal
distribution.

To motivate this two-stage estimator, consider the smoothing dilemma:
If the smoothing level is optimized for MSE, there tends to be too many
$\ell$-change points (wiggles). If the smoothing is chosen to eliminate
the artificial change points (suppress wiggles), then the MSE suffers.
Our two-stage estimate provides the best of both worlds!
In [8], we show that this two-stage estimator achieves the optimal
rate of convergence for functions in the Sobolev space $W_{m,2}$,
while suppressing artificial inflection points in the neighborhood
where they are likely to occur. 
%(provided that $\alpha$ is increased at a specified rate.

In practice, we choose the second-stage smoothing parameter,
$\lambda_{stage \ 2}$, by generalized cross-validation while
%or another data adaptive parameter, 
we scale the first-stage smoothing as
%$h_{stage \ 1}$, to be scaled version of the data adaptive estimate 
$h_{stage \ 1}=\iota(N) h_{GCV}$, 
where $\iota(N) \equiv  log^2(N)N^{\alpha}$ with
$\alpha = \frac{1}{2m +1} - \frac{1}{2\ell+3}$.
Unconstrained smoothing splines can be used in first stage with the
smoothing parameter, $\lambda$, scaled with the correspondence:
$\lambda = h_{eff}^{2m}$, with $h_{eff} \sim (\log N)^2 N^{{-1\over
2\ell +3}}$. 

When $\ell = m-1$, the stage 2 minimization 
%is most easily implemented In this case, we impose 
%linear inequality constraints on $g^{(m)} (t)$ and the 
reduces to a finite dimensional convex  minimization  in the dual formulation. 
%The stage 2 minimization is most easily implemented for $\ell = m-1$. 
%In this case, we impose linear inequality constraints on $g^{(m)} (t)$ 
%and the minimization is finite dimensional in the dual formulation. 
%The free parameters are the $b_m$ spline coefficients,
%$\alpha_i$ where $\hat{g}^{(m)} (t) =P(\sum_{i=1}^m \alpha_i M_i (t))$
%where $M_i (t)$ are the $B$ spline functions and $P$ is the projection
%operator which makes $P(\sum_{i=1}^N \alpha_i M_i (t))=0$ or
%$\sum_{i=1}^M \alpha_i M_i (t)$ depending on whether or not the
%constraints are satisfied.
A simple implementation is that of Villabos and Wahba (VW) [14], 
who add pointwise
constraints on the sign of $\hat{g}^{(m)} (z_m )$ with $z_m$ chosen
in the constraint regions. The goal is to select $\{ z_j ,j=1,
M \}$ such that the constraints are satisfied everywhere even though they
are imposed only at a finite number of points. An important advantage of
our two-stage estimator is that we impose constraints only in small regions
about $\hat{x}_k$. Since the constraint regions are small, the number of
$z_j$ that are necessary in the VW scheme is small. Thus, our
two-stage estimator can be interpreted as a pilot estimator to determine
where to place the $\{ z_j \}$ in the VW scheme.

The VW implementation of the second stage estimator reduces to a
quadratic programming problem with linear inequality constraints. The
number of constraints is bounded by ($m \cdot \hat{k } $ plus {\it 
\# of data points}
{\it in constraint regions}). The B spline representation gives a banded
structure to the programming problem.
The minimization  may be solved exactly using
``active set'' methods in quadratic programming or may be solved approximately
using projected successive over-relaxation (PSOR) iterations. The PSOR method
uses the band structure of convexity constraints %in $A(w)$ 
while active set programs require
modification to take advantage of the tridiagonal structure. 
In both cases, the  dual formulation is easier to implement
%$$\lambda_0 = {\rm arg}  \min_{\lambda} \parallel \Abf_{\omega}^T \lambda +
%y \parallel^2 \ \ \ {\rm subject \ to} \ \lambda \geq 0 \eqno (3)$$
%where $\Abf_{\omega} $ is the $(N-2) \times N$ constraint matrix whose $i$-th
%row is given by (2). 
%The principal advantage of the dual formulation is that
since positivity constraints are substituted  for 
%easier to implement than the three point difference 
inequality constraints.

%\vspace{.5in}

\Section{Adaptive regression splines}

An alternative to smoothing splines is adaptive regression splines [3].
At each step, a new knot is added to the fit. The number and
locations of the new knots are chosen by minimizing a Loss of Fit
(LoF) function,
$d_N (\hat{f},\{ y_i \} )$. Let $\hat{\sigma}^2$ be a measure of the
average residual error:
$
\hat{\sigma}^2 (\hat{f} ,\{ y_i \} ) = {1\over N\sigma^2 }
\sum_{i=1}^N [y_i -\hat{f} (t_i  )]^2
$,
or its $L_1$ analog. Typical discrepancy functions are
\BEQ
d_N  (\hat{f} ,\{ y_i \} ) =\hat{\sigma}^2 \left/
[1-(\gamma_1 p +m)/N ] \right. \  \
{\rm and} \ \
d^B  (\hat{f} ,\{ y_i \} ) = \hat{\sigma}^2
[1+ (\gamma_2 p +m)
\ln (N)/N] \ \ ,
\NEQ
where $p$ is the number of knots in the fit and $m$ is the spline degree. 
Friedman proposed
$d_N^F$ with a default value of $\gamma_1 =3$, while $d^B$ is the
Bayesian/Schwartz information criterion when $\gamma_2 =1$. 
%We believe that 
For a nested family of models, $\gamma_2 =1$ is appropriate 
while $\gamma_2 =2$ corresponds to  a nonnested family with
$ 2 {N \choose K}$
candidate models at the $k$th level [2].
In very specialized settings in regression theory and time series, it has been
shown that LoF functions like $d_N^F$ are asymptotically
efficient while those like $d_N^B$ are asymptotically consistent.
In other words, 
using $d_N^F$-like criteria will
asymptotically minimize the expected error at the cost of not always
yielding the correct model. In contrast, the Bayesian criteria will
asymptotically yield the correct model at the cost of having a larger
expected error.
\medskip

Our goal is to consistently select the number of convexity change
points and efficiently estimate the model subject to the change point
restrictions. Therefore, we propose the following $new$ LoF function:
\BEQ \label{PCICeq}
PCIC = \sigma^2 (\hat{g} ,\{ y_i \} ) \left[
{1+\gamma_2 K \ln (N) /N \over 1-(\gamma_1 p+m)/N } \right] \,
\NEQ
where $K$ is the number of convexity change points and $p$ is the
number of knots. PCIC stands for Piecewise Convex Information Criterion.
In selecting the positions of the $K$ change points, there are essentially
$2 {N\choose K } $ 
%{\left(\! \! {\scriptfont0\begin{array}{c} N \\ K \end{array}}\! \!\right) $
possible combinations of change point locations if we categorize the
change points by the nearest measurement location. 
Thus, our default values are $\gamma_1 =3$ and $\gamma_2 =2$.

We motivate PCIC:
to add a change point requires an improvement in the residual square error
of $O(\sigma^2 \ln (N) )$, which corresponds to an asymptotically
consistent estimate. If the additional knot does not increase the
number of change points, it will be added if the residual error decreases
by $\gamma_1\sigma^2 $. Presently, PCIC is purely a heuristic
principle. We conjecture that it consistently selects the
number of change points and is asymptotically efficient
within the class of methods which are asymptotically consistent
with regards to convexity change points. 
%A stronger conjecture is that PCIC is asymptotically efficient relative to
%estimators which are given the correct number of change points as
%{\it a priori} knowledge.
%\vspace{.2in}

\Section{Instantaneous Frequency and Time-frequency representations}

Much effort has been devoted to finding joint time-frequency
representations of a signal whose frequency is being slowly modulated
in time. The canonical example is a
``chirp'': $y_i = \cos (at + bt^2 )$, where the instantaneous frequency
is $\omega_0 (t) = a+2bt$. In [7], we proposed using the WKB
(eikonal/geometric optics) representation for signals whose amplitude
and frequency are being slowly modulated in time:
\BEQ \label{IFmodel}
y_t =A(\delta t)\cos \left[ \int_0^t \omega (\delta s)ds \right]
+ \epsilon_t \ \ ,
\NEQ
where $\epsilon_t$ is white noise and $\delta$ is a small parameter.
The characteristic time scale for amplitude and frequency modulation is
$1/\delta$. In [7], we propose estimating $A(\delta t)$ and $\omega
(\delta t)$  using data adaptive kernel smoothers. The instantaneous
frequency corresponds to a first derivative estimate. 
We represent the signal in the time-frequency plane as the curve,
$\hat{A}(t, \hat{\omega}(t))$.
Since we expect the phase to be piecewise convex, we replace the adaptive
kernel estimate of $\cos [\phi (t) ]$ 
with PC fitting. % the methods of Sections II and III.
The  circular statistics require  a penalized likelihood functional
of the form:
\BEQ \label{IFmin}
\sum_{i=1}^N \left|{y_i\over \Ah(t_i)} - 
\cos \left[ \int_0^{t_i} \omh ( s)ds \right]
\right|^2
\ + \
\lambda \int |\Ah'''(t)|^2 + |\omh''(t)|^2dt \ \ .
\NEQ
%In [ ], we
%propose an iterative procedure to demodulate the signal and minimize
%the effects of circular statistics. To estimate $\omega (\delta t)$,
%we multiply the signal by $\exp (-i\hat{\omega} t)$: $z_t =y_t \exp
%(-i\hat{\omega} t)$ where $\hat{\omega}$ is the previous  estimate of
%$\omega (\delta t)$. We then kernel smooth the real and imaginary parts of
%$\exp [i\phi (t)]=z_t/|z_t |$.
 Eq.~(\ref{IFmin}) may also
be used in for a smoothing spline fit without PC constraints.
The first term in Eq.~(\ref{IFmin}) differs from Katkovnik [4]
by placing $A(t)$ in the denominator instead of the numerator.

\Section{Spectral Estimation}

Consider a stationary time series, $\{ x_t \}$, with an unknown spectral
density, $S(f)$. A standard method to estimate $S(f)$ is to multiply the
data by a spectral window, compute the windowed periodogram, and then
smooth the windowed periodogram or its logarithm with a data-adaptive kernel
smoother or smoothing splines. In [11-13], we show that a better estimate
replaces the single spectral window with a family of orthonormal spectral
windows. The multiple window estimate reduces the broad band bias while
making a variance stabilizing transformation of the log-periodogram.
When we use the sinusoidal tapers  of  [11],
$\displaystyle{ v^{(k)}_t = \sqrt{{2\over N+1}} \sin \left( {kt \over N+1}
\right) }$, the multi-window estimate reduces to
\BEQ
\hat{S}_{MW} (f) = {\Delta\over K } \sum_{k=1}^K |y(f+k\Delta )
-y(f-k\Delta )|^2 \ , \ \ {\rm where}\ \ \  \Delta =1/(2N+2).
\NEQ
From [12], we recommend $K=(N/2)^{8 /15}$.
Instead of kernel smoothing $\ln [\hat{S}_{MW} (f)]$, we now advocate
using the two-stage
piecewise convex fitting procedure. %of Sections II and III.
Piecewise convex fitting should prove particularly advantageous
to spectral estimation because it will suppress the $1/N$ oscillations
that arise  from discrete time sampling.

\Section{Additive Models, Projection Pursuit and MARS}

Many classes of models attempt to fit multi-dimensional functions as
sums of one-dimensional functions. In each case, we can replace the
standard nonparametric estimation methods with piecewise convex fitting.
 Additive growth curve models [9] fits models of the form
$g(t,x_1 \ldots x_m ) = f_0 (t) +\sum_{j=1}^m f_j (t) x_j $, where
the $x_j$ are determined by smooth splines (old method) or PC fitting
(new method). The back fitting algorithm (corresponding to the Gauss-Seidel iteration)
may be used to PC fit the $f_j$ iteratively.
The same remarks apply to projection pursuit [3].
%$$f(x_1 \ldots x_m ) = \sum_{j=1}^J f_j (v_j \cdot x)$$
%where $v_j$ are direction in $R^m$.
In MARS (multivariate adaptive regression splines), 
a sum of products form:
$ %\displaystyle
{ f(x_1 ,x_2 \ldots x_m )}$
$ %\displaystyle
{ = \sum_{j,j^{\prime} =1}^M g_j (x_j )h_j (x_j^{\prime} ) }$
is assumed.
The knots are placed adaptively, in our case using PCIC (\ref{PCICeq}).

\Section{Robust Estimation}

At present, our understanding of the statistics of false inflection
points is limited to Gaussian errors and linear estimators. In practice, it is
often advantageous to replace both the residual errors 
and the penalty function with more robust analogs: \ 
%\BEQ
$
\sum_{i=1}^N |y_i -\hat{g} (t_i )|^{q_1} + \lambda \int
|\hat{g}^{(m)} (t)|^{q_2} dt
$, %\NEQ
where $1\leq q_i \leq 2$. Representation and duality theorems are given
in [8] for $q_j > 1$.
A heuristic scaling shows that the effective halfwidth of the
robustified function satisfies $h_{eff} \sim [\lambda |f^{(m)}
(t)|^{q_2 -2} ]^{1\over 2m}$. The bias error scales as $g^{(m)} (t) h_{eff}^m$
while the
``variance'' is proportional to $1/Nh_{eff}$. The halfwidth that
minimizes the MISE scales as $h_{eff} \sim N^{{-1\over 2m+1}}$, while the
halfwidth to eliminate false change points of $g^{(\ell )}$ with
asymptotic probability one satisfies $h_{cr} N^{{-1\over 2\ell +3}}
\rightarrow\infty$.
The optimal variable halfwidth kernel smoother has a kernel
halfwidth proportional to $|f^{(m)} |^{{-2\over 2m+1}}$. For
$1\leq q_2 \leq 2$, the effective halfwidth of the robustified
function  automatically reduces the halfwidth in regions of large
$|f^{(m)} (t)|$ just like a variable halfwidth smoother. When $q_1 =
q_2 =1$, the problem reduces to a linear programming problem for each
predetermined set of constraints.

\Section{Image Segmentation}

Image segmentation divides a digital picture into similar
regions for further processing. The Mumford-Shah (MS) algorithm 
assumes that the image is piecewise constant, and that the boundaries
of the regions are unknown [5]. 
The region boundaries are determined
by minimizing the sum of the residual square fit error 
plus a penalty term proportional to the length of the boundary.
If the boundary is parameterized as ${\bf x}(s)$, %,y(s))$, 
the penalty term is the total variation of ${\bf x}(s)$.  %the curve.
We suggest modifying the MS algorithm by replacing the total
variation penalty with a piecewise convex constrained fit using a
robustified penalty function such as the $L_1$ integral of the 
boundary curvature. When using the PCIC (\ref{PCICeq}),
we  use the arclength divided by the grid spacing
as a proxy for $N$, the number of data points.

\Section{Response Surface Estimation}

We seek to estimate a smooth function, $g(x_1 ,x_2 )$, given $N$ noisy
measurements. The two dimensional analog of PC fitting is to divide the
plane into regions where the Gaussian curvature (or more simply
$\Delta f$) has a single sign. The boundaries between regions of positive
and negative Gaussian curvature are free boundaries, which we 
require to be PC. In the first stage of the fit, we use a penalty 
function of the form  $\lambda \int |\Delta^{m/2}g|^2$, where we
require that $\lambda_N^{1/2m} >> log(N) N^{-1 \over 2\ell + 4}$.
This scaling is heuristic since the statistics of false zeros of
$\Delta^{\ell/2} \hat{g}$ are unknown,
as is the critical scaling of the smoothing parameter that avoids extra
regions of incorrectly specified curvature.
To answer these issues, a two-dimensional analog of the Cramer-Leadbetter
formula is necessary.

In the second stage, we suggest imposing constraints on the sign of  
$\partial_{normal} \Delta^{\ell/2} \hat{g}$ near the first-stage
convexity boundaries. The stage 2 convexity boundaries, $(x(s),y(s))$, are
free and need to be  fit using a penalty function plus PC constraints
on the curve shape. The numerical implementation appears tricky with
a need for some elliptic analog of front tracking.  
Overall, the two-dimensional problem appears very challenging
from both  theoretical and numerical perspectives. Basic issues such
as replacing $\partial_{normal} \Delta^{\ell/2} \hat{g}$ with
geometric invariants have not been addressed.

\newpage
\Section{Evolutionary Spectra}

A common model of nonstationary stochastic processes is
$x_t = \int_{-\pi}^{\pi} A(\omega ,\delta t)dZ(\omega )$, where
$dZ(\omega )$ is a stochastic process with  independent spectral
increments $E[dZ(\omega )d\bar{Z}(\omega )] = \delta (\omega -
\omega^{\prime} )d\omega$.
The representation is nonunique for Gaussian processes since $A(\omega ,
t)$ corresponds to a square root of the covariance matrix. To resolve the
nonuniqueness, we require that $A(\omega ,\delta t)e^{i\omega t}$
correspond to the Fourier transform of the positive definite square
root of the covariance matrix.
The evolutionary spectrum is $S(\omega ,t)=|A(\omega ,t)|^2$. 

Let $\lambda_f$ be  the characteristic frequency scale length and $\tau$ 
be the characteristic time scale of $A$: $A (f/\lambda_f,t/\tau)$,
with the sampling rate $=1$. In [6], we present an asymptotic
expansion of the mean square error in estimating $S(f/\lambda_f,t/\tau)$.
We begin by evaluating the multi-window analog of the log-spectrogram on
a two-dimensional time frequency lattice. The bias error is minimized
by using a window length of        %{\cal O}(
$N_w \sim\sqrt{\tau/\lambda_f}$.
In [6], we estimate $\hat{S}(\omega ,t)$ using a two dimensional
cross-product kernel smoother on the log-multi window spectrogram. 
The optimal halfwidths, $h_t$ \& $h_f$,
scale as $h_t/h_f \sim \sqrt{\tau/\lambda_f}$
and $h_fh_t \sim (\tau^2\lambda_f^2)^{-1/3}$, where $h_t$ and $h_f$
are the halfwidths in the $t$ and $f$ directions.

We now advocate replacing kernel smoothing with PC fitting 
the log multi-windowed spectrogram. This method should eliminate the
spurious $1/N_w$ oscillations which occur due to the discrete sampling.
%(The analog of kernel smoothing the log-multi-windowed periodogram in
%Sec.~4.) The only difference is that we estimate a two dimensional.

\Section{\bf Wavelet thresholding: a wiggle enhancer}

Wavelet algorithms  for  function estimation offer two advantages:
1) speed, the algorithms are often $O(N \log (N))$, 
%2) adaptivity to the local smoothness of the unknown function; 3) 
2) asymptotic minimax
optimality in a number of decision theoretic settings [2].
The speed  arises from separability:
each wavelet coefficient is estimated separately without
regard to geometric fidelity. 
The asymptotic optimality theory
assumes that the
unknown function is an arbitrary member of a function space, % such as a Besov
which makes function fits with ten or twenty inflection points
as reasonable as fits with no inflection points. 
Essentially, function spaces contain too many
``unphysical'' functions. 
We prefer  the ``common sense prior''
that the function  has only a few inflection points with high probaability.
%is ignored     %in the assumption 
%that the function is an arbitraryfunction in a Besov space. 

Both of these advantages of wavelets disappear when
the more realistic assumption is made that the unknown
function has only a small number of convexity change points.
Wavelet thresholding  has another intrinsic disadvantage when it comes to
geometric fidelity: 
wavelets possess the complete oscillation property [1].
In contrast, B-splines have the antithetical
and valuable property, total positivity.
%; the number of sign changes of $g(t)$ is less than or equal to the 
%number of sign changes of the sequence of B-spline coefficients.

\Section{Summary}

Sections 1 \& 2 describe nonparametric estimation methods that   
seek to exclude spurious oscillations with negligible sacrifice   
of fit quality. We have outlined a number of applications of
our two-stage piecewise convex fitting method.
The PCF methodology can be used to solve inverse problems.

{\vskip-.1in}
\begin{References}

\item{\normalsize \sc
Chui, C.K.,}
{ An Introduction to Wavelets}, Academic Press, New York, 1992.

\item {\normalsize \sc Donoho, D.L., Johnstone, I.M.}, 
Ideal spatial adaption by wavelet shrinkage, {\bf 81} 425-456
{\it Biometrika}, (1995).

%\item{\normalsize \sc Duarte, A.M. and Vanderbei, R.J.,}
%Interior point algorithms for LSAD and LMAD estimation,
%Princeton University Technical Report SOR-94-07, School of Engineering
%and Applied Science, Princeton, NJ 08544.

\item{\normalsize \sc Friedman, J.}, 
Multivariate adaptive regression splines,
{\em Ann.~Stat.} {\bf 19} 1-141 (1991).

\item{\normalsize \sc Katkovnik, V.}, Local polynomial approximation
of time varying frequency,
Univ.~S.~Africa Stat.~Report 94/1 (1994). 

\item{\normalsize \sc Mumford, D., Shah, J.}, 
Optimal approximation by piecewise smooth 
functions and associated variational problems,
{\em Comm. Pure \& Appl.~Math.},
{\bf 42} 577-685 (1989).

\item {\normalsize \sc Riedel, K.S.}
Optimal data-based kernel estimation
for evolutionary spectra,
{\em I.E.E.E.~Trans.~Signal Processing,} {\bf 41},  2439-2447, (1993).

\item {\normalsize \sc Riedel, K.S.}~Optimal kernel estimation 
of the instantaneous frequency,
{\em I.E.E.E.~Trans.~Signal Proc.,} {\bf 42} 2644-2649, (1994).

\item {\normalsize \sc Riedel, K.S.} Two-stage estimation of
piecewise convex functions, this issue; 
Piecewise convex function estimation and model selection, 
{\em Proc. of Approximation Theory VIII} C.~K.~Chui and L.~L. Schumaker (eds.) 
World Scientific Pub. (1995).

\item{\normalsize \sc Riedel, K.S., Imre K.,}
Smoothing spline growth curves with covariates.
{\em Comm.~in Statistics} {\bf 22} 1795-1818, (1993).

\item {\normalsize \sc Riedel, K.S., Sidorenko, A.,}
Function estimation using data adaptive kernel smoothers -
How much smoothing?
{\em Computers in Physics},
{\bf 8} 402-409, (1994).

\item {\normalsize \sc Riedel, K.S., Sidorenko, A.}
Minimum bias multiple taper spectral estimation. 
{\em I.E.E.E.~Trans.~Signal Processing} {\bf 43} 188-195, (1995).

\item {\normalsize \sc Riedel, K.S., Sidorenko, A.}
Adaptive smoothing of the log-spectrum with multi-tapering.
   Submitted. 
%for publication in {\em I.E.E.E.~Trans.~Signal Processing}.

\item {\normalsize \sc Riedel, K.S., Sidorenko, A., Thomson, D.J.}
Spectral estimation of plasma fluctuations I:
Comparison of methods,
{\em Physics of Plasmas} {\bf 1} 485-500, (1994).

%\item{\normalsize \sc Schwarz, G.,}
%``Estimating the dimension of a model'', {\it Annals of
%Statistics}, Vol. 6, No. 2, 461-464, 1978.

\item{\normalsize \sc M.~Villalobos and G.~Wahba}, %G,
Inequality-constrained multivariate smoothing splines with application
to the estimation of posterior probabilities,
{\em J.~Amer.~Stat.~Assoc.}, {\bf 82}   239-248 (1987).

\end{References} 

{\vskip-.05in}
\begin{Address}
Kurt S.~Riedel, 
Courant Institute,  %of Mathematical Sciences,
New York University, New York, NY 10012-1185
\end{Address}

\end{document}